\documentclass{ctextemp_arxiv-ijmpb}

\begin{document}

\markboth{S.H. Wang}
{Instructions for Typing Manuscripts (Paper's
Title)}

%
\catchline{}{}{}{}{}

\title{Periodicity of chaotic trajectories in realizations of finite
computer precisions and its implication in chaos communications}

\author{Shihong Wang$^{1,2}$  Weirong Liu$^{1}$  Huaping Lu$^{1}$  Jinyu Kuang$^{3}$ and Gang Hu$^{1\ast }$\\
$^{\ast }$\footnote{Corresponding author: ganghu@bnu.edu.cn and
shwang@bupt.edu.cn }}

\address{
$^1$Department of Physics, Beijing Normal University, Beijing\\
100875, China\\
$^2$Science school, Beijing University of Posts and Telecommunications,\\
Beijing 100876,China\\
$^3$Department of Electronics, Beijing Normal University, Beijing
100875,China\\}

\maketitle

\begin{history}
\received{11 12 2003}
\revised{Day Month Year}
\end{history}

\begin{abstract}
Fundamental problems of periodicity and transient process to
periodicity of chaotic trajectories in computer realization with
finite computation precision is investigated by taking single and
coupled Logistic maps as examples. Empirical power law relations
of the period and transient iterations with the computation
precisions and the sizes of coupled systems are obtained. For each
computation we always find, by randomly choosing initial
conditions, a single dominant periodic trajectory which is
realized with major portion of probability. These understandings
are useful for possible applications of chaos, e.g., chaotic
cryptography in secure communication.
\end{abstract}


\section{introduction}    

Chaos study has attracted much attention in the last half of the
20th century [1-7]. However, there have been still some
fundamental problems in the chaos field remaining not completely
understood, including the computer realization of chaos. A
computer has always finite precision, and trajectories of any
autonomous or periodically driven deterministic systems must be
periodic. Thus, we are faced with a difficult circle: chaos, which
generally cannot be analytically shown, has to be manifested by
numerical computations while any computer simulation can never
produce a true chaotic orbit.

In early time of chaos study, the commonly accepted way out of
this difficulty is that the average period of computer realization
of chaotic trajectories(CRCTs) is about the order of the digital
space determined by the computer precision. For high precision
(like double precision $2^{-52}$) this period is very large (may
be even unreachable in numerical simulations), and chaos can be
clearly seen in the time region much smaller than this period.
Moreover, it is anticipated that from different initial conditions
chaotic trajectories of a given dynamics system may eventually
enter different periodic orbits, and the number of periods may be
huge since there are infinitely many unstable periodic orbits
embedded in any given chaotic attractors.

However, the above intuitive understandings are not correct. In
[8-11], the authors showed that the average length of the period
with computation precision $\varepsilon $ is scaled as
$\varepsilon ^{-\frac{D_{2}}{2}}$ with $D_{2}$ being the
correlation dimension of the given chaotic attractor, which is
much shorter than the possible length allowed by the computer
precision. This short periodicity seriously influences some
applications of chaos. For instance, chaos encryption has become
an attracting topic in the recent ten decades[12-18]. The authors
in [19] suggested to use chaotic trajectories of a single map to
reach extremely high level of security, but it was soon found [20]
that the security is very low because the generated chaotic
trajectories by the computer realization are periodic with rather
short average period.

Therefore, the problem of periodicity of CRCTs has both
theoretical significance and practical importance, deserving
further investigation. In this paper we will study this problem by
taking single and coupled map lattices as our examples. We are
particularly interested in the characteristic features of CRCT
affecting the properties of chaotic cryptograph. In Sec.2 we focus
on the periodicity and the related transient of a single map for
different precisions, and find that both the average periods of
CRCTs and the transients are indeed much shorter than the size of
the digital space of computers in all the precisions tested, and
these observations well agree with the theoretical predictions of
[8,11]. A surprising observation is that for different initial
conditions the number of periodic orbits of CRCTs is very small,
and in all precisions tested one can find a single dominant period
for randomly choosing initial conditions which is realized with a
major portion of probability. In Sec.3 we directly compute the
period length of single logistic and tent maps by double precision
realization. In Sec.4 we investigate the same problem for coupled
systems, and find that the average period and transient iterations
increases exponentially as the system size increases. This
observation is very useful in some practical situations (such as
chaos communications) where long periods are required. And the
existence of dominant period is observed as well for coupled
systems with different sizes. The last section gives brief
discussion and conclusion where the significance of the results
obtained in this paper for the applications of chaotic
cryptography is emphasized.

\section{Periodicity and transient of a single chaotic map in finite
precision computer realization}

We first consider the single logistic map as our example

\begin{equation}
x_{n+1}=4x_{n}(1-x_{n})
\end{equation}%
which is well known chaotic, and has aperiodic trajectory for an
arbitrarily chosen typical initial $x_{0}$ in $(0,1)$ region.
However, with finite-precision all trajectories must be periodic,
and the period of the finally realized motion depends on the
initial condition. Our task in this section is to study how the
average period and the average length of transient iterations
towards periodic motions are related to the computer precision;
and how many such periods can be found. All numerical
investigations use Intel Pentium III 700MHz and Pentium IV 1.7GHz
computers with Compaq Visual Fortran code.

Let us follow the computer computation process with round-off
truncations in finite precision. Suppose we take $\varepsilon
=10^{-h}$ $(h=1,2,...)$ computation precision, then for any given
continuous $x_{n}$ value we
measure only discrete value $\stackrel{\wedge }{x}_{n}$, with $\stackrel{%
\wedge }{x}_{n}$ being an integer multiplier of $\varepsilon $ as

\begin{eqnarray}
\stackrel{\wedge }{x}_{n} &=&P(x_{n})  \\
\ \left| x_{n}-P(x_{n})\right|  &\leq &\varepsilon /2  \nonumber
\end{eqnarray}%
With (2) the actual map becomes

\begin{eqnarray}
x_{n+1} &=&4\stackrel{\wedge }{x}_{n}(1-\stackrel{\wedge }{x}_{n})
\\
\stackrel{\wedge }{x}_{n} &=&P(x_{n})  \nonumber
\end{eqnarray}%
In Eqs.(3) with the discrete operation the state $x_{n}=0$ may be
realized with certain finite probability [in fact this probability
is zero for
continuous variables, i.e., $x_{n}=0$ cannot be realized unless $x_{0}\ $%
takes certain nontypical values of zero measures]. Throughout the
paper, we will exclude the $x_{n}=0$ state solution from our
counting. Map (3) is always periodic though map (1) is chaotic and
aperiodic. As an example we consider $h=4$, and start from
$x_{0}=0.951636985290801$. The iterations of (3) give
\begin{eqnarray}
x_0 &=&0.951636985290801\ (\stackrel{\wedge
}{x}_{0}=0.9516)\rightarrow x_1=0.18422976\ (\stackrel{\wedge
}{x}_1=0.1842)\rightarrow \cdots \rightarrow  \nonumber
\\ x_{14} &=&0.453879\
(\stackrel{\wedge }{x}_{14}=0.4539)\rightarrow x_{15}=0.99149916\
(\stackrel{\wedge }{x}_{15}=0.9915)\
\rightarrow \cdots \rightarrow   \\
x_{117} &=&0.54599356\ (\stackrel{\wedge
}{x}_{117}=0.5460)\rightarrow x_{118}=0.991536\ (\stackrel{\wedge
}{x}_{118}=0.9915) \nonumber
\end{eqnarray}
For this initial condition the period of map (3) is
$T=118-15=103$, and the
transient time for the trajectory to enter the periodic circle is $\tau =15$%
. Of course, for different $x_0$'s we may find different periods
and different lengths of transient iterations.

With the above approach we can systematically investigate the
dependence of $\tau $, $T$ on the precision $\varepsilon
=10^{-h}$. In Fig.1 we plot $\tau $ (circle) and $T$ (square) vs
$h$, where $\tau $ and $T$ are obtained as the averages with
arbitrarily chosen different initial $x_0$ values,

\begin{equation}
\tau =\frac{1}{M}\sum\limits_{i=1}^{M}\tau _{i},\ \ T=\frac{1}{M}%
\sum\limits_{i=1}^{M}T_{i}
\end{equation}%
where $\tau _{i}$ and $T_{i}$ are the transient and period
iterations computed by $i$th initial condition, respectively, and
$i$ runs over all possible initial states for $h\leq 6$
($M=10^{h}$) in the discrete variable space, and $i$ is chosen
randomly in $(1,10^{h})$ for $h>6$. In Fig.1
vertical bars indicate error estimates $\sigma _{\tau }=\sqrt{\frac{1}{M}%
\sum\limits_{i=1}^{M}(\tau _{i}-\tau )^{2}},\sigma _{T}=\sqrt{\frac{1}{M}%
\sum\limits_{i=1}^{M}(T_{i}-T)^{2}}$. From the data, empirical
exponential forms

\begin{equation}
\tau ,T\ \approx 10^{\alpha +\beta h},\ \ \alpha \approx -0.40,\ \
\beta \approx 0.47
\end{equation}%
can be approximately formulated [the solid curve in Fig.1]. It is
emphasized that the period $T$'s (also the transient time $\tau
$'s) are much shorter than those allowed by the spaces of the
computer data for the given precisions. For instance, we have $T$
$\approx 2\times 10^{3}<<10^{8}$ for $10^{-8}$ precision and $T$
$\approx 10^{6}<<10^{13}$ for $10^{-13}$ precision.

\begin{figure}[1th]
\centerline{\psfig{file=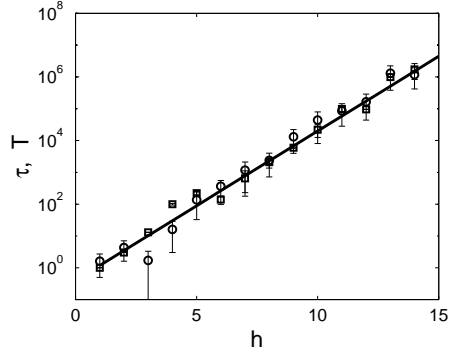,width=6cm}} \vspace*{8pt}
\caption{Average transient time $\tau $'s (circle) and average
period $T$'s (square) for entering periodicity of numerical
realizations of chaotic trajectories of Eq.(3) for different
precisions. Solid line is the result of the empirical formulas
(6), which fits the numerical data satisfactorily. Vertical bars
are errors of statistical data of average $\tau $ and $T$ for
different precisions.}
\end{figure}

  There are several interesting observations in Fig.1 and Eq.(6)
worthwhile remarking. First, the average period length has a power
law scaling against the discrete size $\varepsilon =10^{-h}$, as
$T\ \propto \varepsilon
^{-\beta }$, $\beta \approx -\frac{D_{2}}{2}$(within the statistical error $%
\pm 0.02$) with $D_{2}$ being the correlation dimension of the
given chaotic attractor (in our case $D_{2}=1$), this agrees with
the theoretical prediction of [8,11]. Moreover, the average
transient time $\tau $ has the similar scaling as $T.$ From the
point of view of cryptography, transient time $\tau $ has the same
significance as period $T$ itself. For a given discrete size
$10^{-h}$, there are a number of periodic orbits of which the
corresponding periods vary largely. There exist some orbits with
very small periods even $h$ is relatively large. Large $\tau $ is
useful for keeping the communication security even in the case of
small $T$, since it is not the quantity $T$ but the quantity
$T+\tau $ that is related to the security problem of the chaotic
cryptography.

Another interesting feature of CRCT is that the number of periodic
orbits which the system eventually enters for a given computation
precision is very small. An intuitive picture is: Since there are
infinitely many unstable periodic orbits densely embedded in any
chaotic attractor it is easy for a trajectory to be trapped in one
of these orbits after a course grain caused by finite precision
computation, and thus the number of realized periodic orbits may
be huge as the computation precision is high $(h\gg 1)$. To our
great surprise, for any precisions (up to the precision $10^{-14}$
) the numbers of observed distinctive orbits are very small, and
for most of truncation precisions we observe that one orbit of all
periods takes major part of probability, and dominates the
statistical behavior of CRCT. These results are listed in Table 1,
which are strangely against the above intuition. Note, the
existence of such a dominant period dose not support the
theoretical results on the probability of periods in [8,10]. The
reason for this disagreement may be the following. In [8] the
prediction is proven based on approximating the map to be
deterministic in time and purely random in variable space map
(though deterministic [8,10]). The existence of dominant period
breaks the validity of this intuitive picture.
\begin{table}[h]
\tbl{Period lengths and corresponding probabilities by different
decimal precisions realization.}
 {\begin{tabular}{@{}cccc@{}} \toprule
 Precision & Number & Period length $T_i$ (ratio) \\
$\varepsilon =10^{-h}$  & of tests \\ \colrule
$10^{-1}$\hphantom{00} & \hphantom{0}$10^{1}$ & \hphantom{0}1(1.00) \\
$10^{-2}$\hphantom{00} & \hphantom{0}$10^{2}$ & \hphantom{0}3(0.60) 1(0.40)\\
$10^{-3}$\hphantom{00} & \hphantom{0}$10^{3}$ & \hphantom{0}1(0.92) 13(0.08)\\
$10^{-4}$\hphantom{00} & \hphantom{0}$10^{4}$ & \hphantom{0}103(0.4916) 97(0.3298) 1(0.1760) 6(0.0014) 2(0.0012) \\
$10^{-5}$\hphantom{00} & \hphantom{0}$10^{5}$ & \hphantom{0}227(0.92279) 1(0.06093) 29(0.00937)\\
& & 23(0.00661) 5(0.0016) 4(0.00008) 2(0.00006)\\
$10^{-6}$\hphantom{00} & \hphantom{0}$10^{6}$ & \hphantom{0}155(0.680588) 1(0.210002) 16(0.065585) 88(0.033479)\\
& & 79(0.00826) 8(0.001122) 30(0.000512) 11(0.000336) 6(0.000116)\\
$10^{-7}$\hphantom{00} & \hphantom{0}$10^{3}$ & \hphantom{0}1078(0.505) 136(0.427) 1(0.068)\\
$10^{-8}$\hphantom{00} & \hphantom{0}$10^{3}$ & \hphantom{0}2412(0.756) 419(0.122) 1(0.10) 428(0.018) 118(0.004)\\
$10^{-9}$\hphantom{00} & \hphantom{0}$10^{3}$ & \hphantom{0}5957(0.97) 1(0.028) 860(0.002) \\
$10^{-10}$\hphantom{00} & \hphantom{0}500 & \hphantom{0}26358(0.814) 1319(0.132) 1680(0.026) 4317(0.012) \\
& & 1(0.012) 1233(0.002) 643(0.002)\\
$10^{-11}$\hphantom{00} & \hphantom{0}200 & \hphantom{0}101487(0.96) 22916(0.035) 2982(0.005)  \\
$10^{-12}$\hphantom{00} & \hphantom{0}200 & \hphantom{0}1(0.815) 95914(0.185) \\
$10^{-13}$\hphantom{00} & \hphantom{0}66 & \hphantom{0}1(0.818) 985982(0.182) \\
$10^{-14}$\hphantom{00} & \hphantom{0}28 & \hphantom{0}2385320(0.429) 1908178(0.357) 183458(0.107) 1(0.107) \\
\botrule
\end{tabular}}
\end{table}

 In the above discussion we used the decimal representation and the
fixed-precision truncations. The actual computer round-off
discretization use the binary representation and
relative-precision truncations. In Fig.2 we do the same as Fig.1
by applying binary representation and relative-precision
truncations. The results are similar to those of Fig.1, showing
that though the discretization of variable space has some
significant changes in the dynamics and statistics of chaotic
systems, the detailed methods of discretization is not crucial.

\begin{figure}[2th]
\centerline{\psfig{file=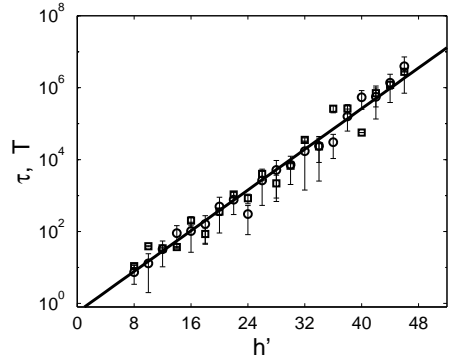,width=6cm}} \vspace*{8pt}
\caption{The same as Fig.1 with binary representation and
relative-precision truncations applied. The statistical results
are unchanged from Fig.1.}
\end{figure}

\section{Periodicity of single chaotic maps in double
precision computer realization}

We directly and numerically compute Eq.(1) in Table 2 by the
double precision $2^{-52}$ computations, and by using the
arithmetic of [20] to detect the period length. This method cannot
exactly give transient time, but can record period exactly and
effectively. A significant advantage of this approach is that it
can compute extremely long period with small storage space and
fast computation. We can see only $12$ different periods for
$10^{5}$ different initial conditions from Table 2. There exists a
major probability period and there are near to zero probabilities
for some periods. The average period length is about $5.7\times
10^{6}$ iterations, which fits form (6) approximately. The results
of the direct double precision computing well confirm our
discretization method. We use C code to validate it. It is
interesting to point out that for the true double precision
computation of the model Eq.(1), the period of length 5638349
(which is one order smaller than the period $T\approx
2^{26}\approx 6\times 10^{7}$ predicted by [8]) can appear with
probability $67.75\%$ ($10^{5}$ different initial conditions are
computed). This knowledge may be useful for scientists interested
in computer simulations of chaotic systems. We also consider the
single tent map $x_{n+1}=1-\left| 1.95x_{n}-1\right| $ as our
example and compute the period length in Table 2 by the double
precision. There are the similar results as the single logistic
map.

\begin{table}[h]
\tbl{Period lengths and corresponding numbers of $10^5$ different
initial conditions of logistic and tent maps by double precision
realization.}
 {\begin{tabular}{@{}cccc@{}cccc@{}cccc@{}} \toprule
 $T_i$ & Numbers & Numbers & Numbers & Numbers & Numbers \\
(Logistic map)  & of $20000$ & of $20000$ & of $20000$ & of $20000$ & of $20000$ \\
\colrule
5638349\hphantom{00} & \hphantom{0}13533 & \hphantom{0}13672 & \hphantom{0}13533 & \hphantom{0}13482 & \hphantom{0}13631\\
1\hphantom{00} & \hphantom{0}3233 & \hphantom{0}3312 & \hphantom{0}3235 & \hphantom{0}3313 & \hphantom{0}3253\\
14632801\hphantom{00} & \hphantom{0}2339 & \hphantom{0}2297 & \hphantom{0}2284 & \hphantom{0}2272 & \hphantom{0}2172\\
10210156\hphantom{00} & \hphantom{0}336 & \hphantom{0}346 & \hphantom{0}349 & \hphantom{0}346 & \hphantom{0}357\\
2441806\hphantom{00} & \hphantom{0}284 & \hphantom{0}290 & \hphantom{0}302 & \hphantom{0}285 & \hphantom{0}275\\
2625633\hphantom{00} & \hphantom{0}264 & \hphantom{0}274 & \hphantom{0}285 & \hphantom{0}290 & \hphantom{0}301\\
420909\hphantom{00} & \hphantom{0}7 & \hphantom{0}1 & \hphantom{0}3 & \hphantom{0}3 & \hphantom{0}3\\
960057\hphantom{00} & \hphantom{0}4 & \hphantom{0}3 & \hphantom{0}5 & \hphantom{0}5 & \hphantom{0}6\\
1311627\hphantom{00} & \hphantom{0}0 & \hphantom{0}4 & \hphantom{0}2 & \hphantom{0}2 & \hphantom{0}2\\
4389\hphantom{00} & \hphantom{0}0 & \hphantom{0}1 & \hphantom{0}1 & \hphantom{0}0 & \hphantom{0}0\\
510250\hphantom{00} & \hphantom{0}0 & \hphantom{0}0 & \hphantom{0}1 & \hphantom{0}1 & \hphantom{0}0\\
234209\hphantom{00} & \hphantom{0}0 & \hphantom{0}0 & \hphantom{0}0 & \hphantom{0}1 & \hphantom{0}0\\
\colrule
$T_i$ & Numbers & Numbers & Numbers & Numbers & Numbers \\
(Tent map)  & of $20000$ & of $20000$ & of $20000$ & of $20000$ & of $20000$ \\
\colrule
216135802\hphantom{00} & \hphantom{0}10111 & \hphantom{0}10105 & \hphantom{0}10089 & \hphantom{0}10045 & \hphantom{0}10053\\
1009008\hphantom{00} & \hphantom{0}7965 & \hphantom{0}7966 & \hphantom{0}7968 & \hphantom{0}8006 & \hphantom{0}7993\\
128662357\hphantom{00} & \hphantom{0}1820 & \hphantom{0}1824 & \hphantom{0}1836 & \hphantom{0}1847 & \hphantom{0}1851\\
8703769\hphantom{00} & \hphantom{0}47 & \hphantom{0}48 & \hphantom{0}49 & \hphantom{0}45 & \hphantom{0}44\\
15083761\hphantom{00} & \hphantom{0}35 & \hphantom{0}34 & \hphantom{0}36 & \hphantom{0}35 & \hphantom{0}38\\
4387781\hphantom{00} & \hphantom{0}15 & \hphantom{0}16 & \hphantom{0}15 & \hphantom{0}15 & \hphantom{0}15\\
3404560\hphantom{00} & \hphantom{0}6 & \hphantom{0}6 & \hphantom{0}6 & \hphantom{0}6 & \hphantom{0}6\\
1299622\hphantom{00} & \hphantom{0}1 & \hphantom{0}1 & \hphantom{0}1 & \hphantom{0}1 & \hphantom{0}0\\

\botrule
\end{tabular}}
\end{table}

\section{Periodicity of chaotic coupled map lattices in Computer realization}

In Figs.1 and 2 it is clearly observed that with a single chaotic
map the period of orbits of computer realization is rather short.
For double precision ($2^{-52}$) the average period is about of
order $10^{6\sim 7}$, which can be very easily reached. This short
period problem may cause serious disadvantages in chaos
applications. For instance, a single autonomous chaotic map can
never allow high security in computer realization of chaotic
cryptography since long period is a necessary condition for any
kinds of cryptosystems [19-21]. In Ref.22, the authors suggested
to use spatiotemporal chaotic systems for cryptography and
declared that they can reach very high level of security, it is
then interesting to investigate the periodicity of CRCTs of such
spatiotemporal systems.

Let us extend the single map (1) to one-way coupled map lattice

\begin{eqnarray}
x_{n+1}(i) &=&(1-\varepsilon )f[x_{n}(i)]+\varepsilon
f[x_{n}(i-1)],\ \
i=1,2,...,N   \\
f(x) &=&4x(1-x),\ \ \ x(0)=x(N)  \nonumber
\end{eqnarray}%
where we use periodic boundary condition of system size $N$ , and take $%
\varepsilon =0.95$. This system is used for chaotic encryption in
[22].

We are now interested in how both the system size $N$ and the
computer precision $10^{-h}$ affect the period $T$ and the
transient time $\tau $ of CRCTs. In Figs.3(a)-(c) we do the same
as Fig.1 with $N=2$, $N=3$ and $N=4$ in Eq.(7), respectively. The
average lengths of period and the transient iterations increase
exponentially with $h$ much faster than Fig.1. Moreover, for the
same precision, the average lengths of the period and the
transient are longer if the system size is taken larger. In
Fig.3(d) we fix $h=3$ and plot $T$ and $\tau $ vs $N,$ an
exponential increasing of $T$ and $\tau $ with $N$ are confirmed.
In all figures of Fig.1 and Fig.3, the solid curves are drawn from
the united empirical form

\begin{equation}
T_{N}^{(h)},\tau _{N}^{(h)}\hspace{0in}\hspace{0in}\ \propto \
10^{(\alpha +\beta h)N}=10^{\alpha N}\varepsilon ^{-\beta N},\ \
\alpha \approx -0.40,\ \ \beta \approx 0.47
\end{equation}%
which coincides satisfactorily with the numerical data for wide
range of $N$ and $h$. By increasing the system size $N,$ the
coupled systems can manifest its chaoticity in much longer time
though periodicity must appear for sufficiently large time scale.
If the system size is sufficiently large the periodic behavior of
CRCTs is practically not observable. According to the
empirical formula (8), for $N=4$, and double precision $\varepsilon =2^{-52}$%
, the periodicity occurs after $10^{28}$ iterations, and a common
PC (with 1.7G CPU) needs $10^{12}$ years, to make so many
iterations. Thus, the short period problem of CRCTs can be
practically solved by applying coupled (spatiotemporal) chaotic
systems with a sufficiently large $N$. This is one of several most
important reasons why the system in [22] may reach very high
security.
\begin{figure}[3th]
\centerline{\psfig{file=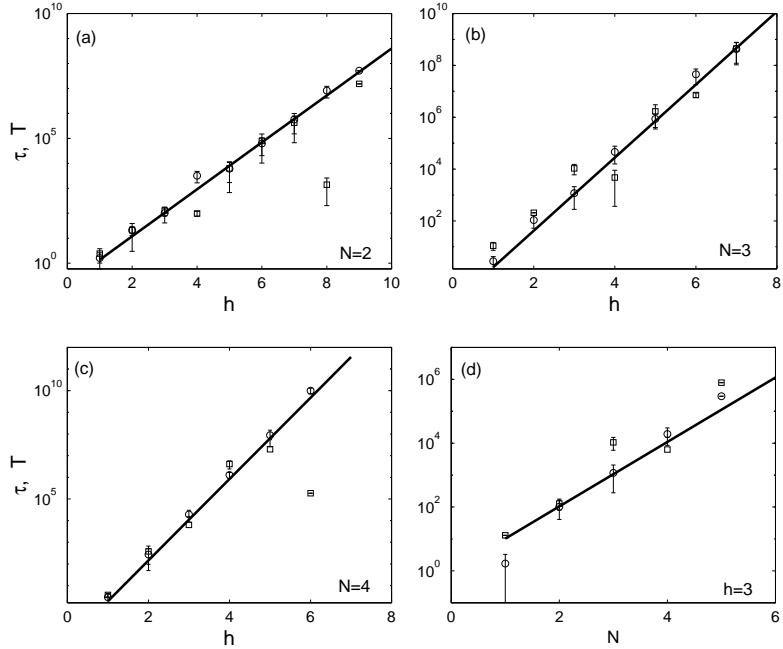,width=11cm}} \vspace*{8pt}
\caption{(a)-(c) The same as Fig.1 with Eq.(7) considered.
$\varepsilon =0.95,$ $N=2$, $N=3$, and $N=4$, respectively. (d)
$T$ and $\tau $ plotted vs $N$ for Eq.(7) with $h=3$. Solid lines
are the results of the empirical formulas (8), which fit well the
numerical data. It is shown that the average transient $\tau $
following the power law Eq.(8) satisfactorily with smaller
fluctuation than that of $T$.}
\end{figure}

The phenomenon of small number of periodic orbits of CRCT is also
observed for coupled maps. In Fig.3 for $N>1$ and relatively large
$h$, we can try only small number of tests for each given
precision $h$. Therefore, the fluctuation of $T$ is rather larger.
Sometimes we obtain very small $T$'s, which deviate from the
exponential law (the solid line of Fig.3) considerably. However,
almost in all our tests the transient time $\tau $'s
follow Eq.(8) satisfactorily with fluctuation much smaller than that of $T$%
's (see data of Fig.3). Since the cryptography of chaotic systems
is related to $T+\tau $ rather than $T$, such nice power law
behavior of $\tau $ is useful for controlling the security of
cryptosystems.

\section{discussion and conclusion}

In conclusion we have investigated the problems of periodicity and
transient to periodicity of chaotic trajectories in computer
realization with finite computation precision. It is found that
both average period and the transient time to periodic orbits have
power law relation with the computer round-off precision. For
low-dimensional systems the periods of chaotic trajectories and
the corresponding transients can be rather short even the
double-precision computation is applied. And for each precision
the number of periodic orbits of computer realization is small
even if the precision is rather high and the number of tests is
very large. The problems of short period and small number of
periodic orbits may seriously affect the applications of chaos,
e.g., it results in weak cryptography in secure communication.
Nevertheless, this problem can be satisfactorily solved by
applying spatiotemporal chaos (or say, coupled chaotic systems).
With few coupled chaotic subunits the period of CRCTs may be as
large as practically unreachable. A surprising observation
worthwhile remarking is that by applying different initial
conditions there always exists a dominant period (for all tested
computation precisions and sizes of coupled systems) which appears
with major portion of probability.

This work was supported by the National Science Foundation of
China and Nonlinear Science Project.

\end{document}